\documentclass[fleqn,usenatbib]{mnras}
\usepackage{newtxtext,newtxmath}

\usepackage[T1]{fontenc}

\DeclareRobustCommand{\VAN}[3]{#2}
\let\VANthebibliography\thebibliography
\def\thebibliography{\DeclareRobustCommand{\VAN}[3]{##3}\VANthebibliography}

\usepackage{hyperref} 
\usepackage{graphicx}
\usepackage{graphicx}	
\usepackage{amsmath}	
\usepackage{amssymb}	
\usepackage[pstricks]{bclogo}

\allowdisplaybreaks

 \hypersetup{
	pdftitle={Influence of DM on gravitational stability of isotermal gas clouds},
	pdfsubject={Astrophysics...},
	pdfauthor={Kalashnikov I.},
	pdfkeywords={DM... }
}
\graphicspath{{figs/}}
\title[Influence of dark matter on gravitational stability of isothermal gas clouds]{Influence of dark matter on gravitational stability of isothermal gas clouds}

\author[Kalashnikov, Chechetkin]{
I. Yu. Kalashnikov,$^{1}$\thanks{E-mail: kalasxel@gmail.com}
V. M. Chechetkin,$^{1,2,3}$
\\
$^{1}$Keldysh Institute of Applied Mathematics, 4 Miusskaya sq., Moscow, 125047, Russia\\
$^{2}$National Research Center ''Kurchatov Institute'', 1 Akademika Kurchatova sq., Moscow, 123182, Russia\\
$^{3}$National Research Nuclear University MEPhI, 31 Kashirskoe sh., Moscow, 115409, Russia
}

\date{Accepted XXX. Received YYY; in original form ZZZ}

\pubyear{2022}

\begin{document}
\label{firstpage}
\pagerange{\pageref{firstpage}--\pageref{lastpage}}
\maketitle

\begin{abstract}
To date, the presence of dark matter (DM) can be judged only by its gravitational interaction on the visible matter. It is therefore important to find the consequences of this interaction, which can then help to determine both the DM properties and parameters and the dynamics and evolution of visible matter. The gravitational influence of dark matter on the stability of interstellar medium (ISM), the progenitor of stars and star clusters, was considered. An isothermal self-gravity gas was taken as a suitable model describing ISM, particles interacting only gravitationally were considered as DM. The results obtained by analytical methods show that even a small amount of fast DM particles significantly increases the stable radius of the gas cloud and the corresponding mass while a higher relative density of DM destabilizes the gas. It was shown that with typical parameters of ISM and DM, its presence increases the maximum stable mass of isothermal cloud by a factor of four and the radius by five.
\end{abstract}



\section{Introduction}\label{sec:Intro}

Since obtaining of the first observational evidences to this day, the nature of dark~matter~(DM) remains one of the most mysterious questions of the modern astrophysics. Intensive research in this area (\cite{Bertone18}) has led to currently the most common view that DM consists of neutral stable particles with a small cross-section of an interaction with particles of the visible baryonic~matter~(BM). There is usually no reason to introduce a DM particle interaction other than gravitational one. Thus, when talking about observational manifestations of DM, one is usually referring to its gravitational influence on BM, such as the non-Keplerian rotation curves of galaxies (\cite{Rubin1970,Bosma1978,Rubin1980}), the dynamics of galaxy clusters (\cite{Zwicky37,Allen11}) and gravitational lensing (\cite{Clowe2004}). It is worth mentioning the so far unsuccessful attempts both to detect dark matter directly (\cite{Undagoitia15}) and to construct alternative theories of gravity (\cite{Milgrom83,Bekenstein04}). 
Since so far the presence of DM is known only through the gravitational interaction, it is important to find other consequences of this interaction, which may further help in determining the DM properties and parameters and also to find out how the behaviour of BM changes in the presence of DM with parameters determined by other methods. In this paper, we considered a gravitational effect of DM presence on the stability of interstellar gas, the progenitor of stars and star clusters. 

The consideration of self-gravitating gas stability problems was started by \cite{Jeans1902}, who studied homogeneous state of gas at rest. Later on this theory was extended to the case of an expanding universe (\cite{Bonnor57}), taking into account the effects of general relativity (\cite{Lifshitz46}). The unquenchable interest in this problem is caused by open questions about the formation of the large-scale structure of the universe (\cite{Shandarin83,Longair08,Demianski11}). Investigation of kinetic theory of Jeans instability were initiated by \cite{Lynden-Bell62} and \cite{Sweet62} and it continues to be developed (\cite{Trigger04,Rozina17,Yang20,Kremer21}). Besides, studies were carried out taking into account the presence of turbulence (\cite{Chandrasekhar51,Bonazzola87}), magnetic fields (\cite{Parker66,Kumar90}) and dust (\cite{Pandey94,Rozina17}). 

In his consideration Jeans did not take into account thermodynamic state of the self-gravitating gas, holding its density as a constant. However, such a stationary 
configuration cannot exist (see Section~\ref{sec:swindle}). Easy-to-analyze and at the same time sufficiently accurate models of a self-gravitating gas have been known for a long time. They have barotropic equation of state $p=K \rho^\gamma$, where $p$, $\rho$ are pressure and density, $K$, $\gamma$ are free parameters. The case $\gamma>1$ corresponds to the polytropic models widely used in the study of stellar evolution (\cite{Chandrasekhar39}). If $\gamma=1$ and $K=c_s^2$, where $c_s$ is speed of sound, the equation describes an isothermal gas, which, when self-gravity is taken into account, gives a description of a cloud of gas, commonly used as a model for ISM -- interstellar medium (\cite{Draine11}). The results of the stability investigation of this configuration, done by \cite{Ebert55,Bonnor56}, showed only a slight difference from Jeans theory. Subsequently, these studies were continued mainly in connection with the general stability properties of self-gravity systems and the development of the corresponding methods for their investigation (\cite{Antonov62,LyndenBell68,Katz78,Padmanabhan89,Chavanis02b,Chavanis02c}). 


In the early days of the search for DM particle candidates, neutrinos were considered (\cite{Lee77,Zeldovich80}). However, further studies of the formation of large-scale structures of the universe have shown the failure of this assumption (\cite{White83}).
After that, researchers' attention shifted to other possible candidates, the most discussed of which are hypothetical weakly interacting massive particles -- WIMPs (\cite{STEIGMAN85,Cerdeno2009}). In this work the effects of the presence of such particles on the stability of isothermal self-gravity gas as an appropriate model describing ISM are investigated. In doing so, our consideration is not limited to WIMPs, but can also be applied to other candidates whose behavior can be considered statistically in a limited area without self-interaction other than gravitational one.

\section{preview}

\subsection{Isothermal gas spheres}\label{sec:unprtbdIsoterm}

To begin with let us shortly consider the theory of isothermal gas spheres and their stability. The problem of finding the equilibrium configuration of such a sphere is reduced to finding the gravitational potential $\phi$ with a density $\rho$ known from the Boltzmann distribution:
\begin{align}
	& \frac{1}{r^2}\frac{d}{dr}r^2\frac{d\phi}{dr} =4\pi G \rho, \\
	& \rho = \rho_c \exp\left( -\frac{m_b \phi}{kT} \right), \label{bolzDist}\\
	& \phi|_{r=0} = \frac{d\phi}{dr}\bigg|_{r=0} = 0,
\end{align}
where the first boundary condition means that the potential is counted from zero at the center, the second one corresponds to zero gravity at the center of the sphere; $\rho_c$ is density in the center, 
$m_b$ is a mean mass of BM particles, $T$ is a constant temperature of the gas, $G$, $k$ -- gravitational and Boltzmann constants respectively. Introducing the notations $\psi = m_b \phi /kT$, $\zeta = r/r_0 = r\sqrt{4\pi G\rho_c m_b /kT}$ we come to the following problem:
\begin{align}
&	\frac{d^2\psi}{d\zeta^2} +\frac{2}{\zeta} \frac{d\psi}{d\zeta} = e^{-\psi}, \label{psiEQ}\\
&	\psi|_{\zeta=0} = \frac{d\psi}{d\zeta}\bigg|_{\zeta=0} = 0. \label{psiEQbnd}
\end{align}

The mass $M$ enclosed in a sphere of radius $r$ is the integral of $4\pi r^2\rho$, which may be rewritten using (\ref{psiEQ}) in the form:
\begin{align}
	M(r) = M(r_0 \zeta)= 4\pi\rho_c r_0^3 \zeta^2\frac{d\psi}{d\zeta}.
	\label{massSphere}
\end{align}
Next, we may express the central density in terms of the whole mass $M_\text{tot}$ of the considering sphere with radius $R=r_0 \varkappa$:
\begin{align}
	 \rho_c = \frac{\mu_s^2}{4\pi M_\text{tot}^2}\left( \frac{kT}{m_b G} \right)^3,
	 \label{centDens}
\end{align}
where $\mu_s = \varkappa^2 \psi'(\varkappa)$ is the dimensionless sphere mass, $\varkappa$ -- dimensionless radius of the sphere, the prime denotes $\zeta$ derivative.

The gas pressure on the surface of the sphere should be balanced by the external pressure $P_s$. Taking into account the ideal equation of state and (\ref{bolzDist})(\ref{centDens}) relations we have:
\begin{align}
	P_s = \frac{1}{4\pi G^3 M_\text{tot}^2}\left( \frac{kT}{m_b} \right)^4 \cdot \varkappa^4 \psi'^2(\varkappa) e^{-\psi(\varkappa)} \stackrel{\text{def}}{=} \Pi_s \cdot \pi_s,
	\label{surfPres}
\end{align} 
where the dimensionless pressure $\pi_s = \varkappa^4 \psi'^2(\varkappa) e^{-\psi(\varkappa)}$ was introduced. Thus, setting $m_b$, $M_\text{tot}$, $T$ and the external pressure~$P_s$ we 
may find out the radius of such gas sphere with solving $P_s/\Pi_s(M_\text{tot},T,m_b) = \pi_s(\varkappa)$ equation relatively $\varkappa$, thereby to reveal the remaining characteristics using above relations.

\begin{figure}
	\centering
	\includegraphics[width=1\linewidth]{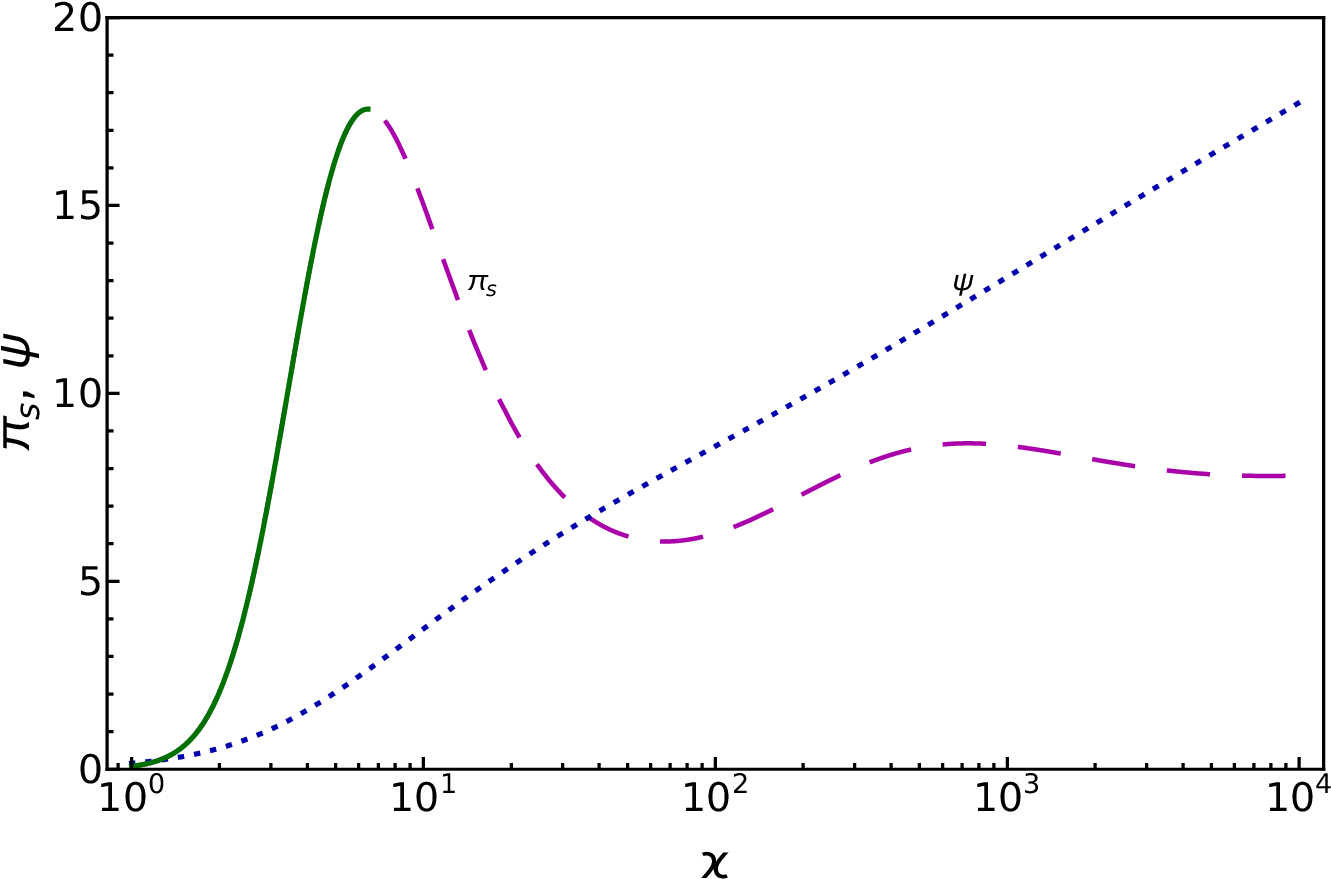}
	\caption{Dependence of the dimensionless external pressure $\pi_s$ on the dimensionless radius $\varkappa$ of an isothermal sphere. Stable configurations correspond to the solid line, unstable -- dashed. The potential $\psi$ is represented via dotted line. }
	\label{fig:pressStab}
\end{figure}

The dependence $\pi_s(\varkappa)$ is shown on Figure~\ref{fig:pressStab}, according to which the dimensionless pressure does not increase monotonically with increasing of sphere radius. As it was shown by \cite{Ebert55,Bonnor56}, the global maximum $\varkappa^\text{max}=6.45$, $\pi_s^\text{max}=17.65$ separates regions of stable and unstable configurations. If the external pressure is higher than this critical value $P_s>\Pi_s \pi_s^\text{max}$ then the elasticity of the gas is insufficient to withstand it and compression must occur. The critical mass corresponding to this pressure is called the Bonnor - Ebert mass, it may be gotten from the definition (\ref{surfPres}) of $\Pi_s$:
\begin{align}
	M_\text{tot}^\text{max} = \frac{\alpha}{\sqrt{P_s G^3}} \left( \frac{kT}{m_b} \right)^2, \label{BEmass}
\end{align}
where $\alpha=\mu_s e^{-\psi(\varkappa)/2}/\sqrt{4\pi}=1.18$. The radius of such a sphere may be expressed as:
\begin{align}
	R^\text{max} = \frac{\beta}{\sqrt{P_s G}}  \frac{kT}{m_b} , \label{BErad}
\end{align}
with $\beta=\varkappa e^{-\psi(\varkappa)/2}/\sqrt{4\pi}=0.49$, which has nonmonotonic dependence on $\varkappa$.
Further consideration is devoted to the question of the influence of DM on the Bonnor - Ebert mass and the corresponding radius. 

\subsection{Jeans swindle}\label{sec:swindle}
The gravitational stability of a homogeneous medium at rest was considered by \cite{Jeans1902}. The initial prerequisites for such a consideration were not entirely correct, since within the framework of Newtonian theory, there is no equilibrium stationary distribution of matter. The fact is that from the Euler equation at zero velocities and pressure gradients it follows that the gravitational potential must be constant, while according to the Poisson equation this is possible only in the absence of matter. Nevertheless, consideration of the perturbations propagation against the background of a constant gravitational potential (i.e.\@ zero unperturbed gravitational force) leads to the correct results. The assumption that the unperturbed gravitational field is inessential for the propagation of perturbations is called the Jeans swindle.

A correct stability analysis of a uniformly distributed gas against the background of an expanding universe was made by \cite{Bonnor57}. Before this, a similar problem within the framework of the general relativity was considered by \cite{Lifshitz46}. These investigations showed an almost exact match with critical mass of a gas obtained by \cite{Jeans1902}. Future consideration (\cite{Kiessling03,Falco13}) showed that  Jeans swindle is not a trick, but a completely physically grounded assumption, which at one time could not appear in a self-consistent form due to the absence of the general theory of relativity. In this work, we proceeded from Newton's theory of gravity and the uniform distribution of DM, assuming the validity of Jeans swindle because of the reasons described above.

\section{Problem formulation}\label{sec:Probl}

\subsection{Governing equations}\label{sec:govEqs}
We continued to describe the motion of a fluid hydrodynamically, while the motion of DM particles was described using the collisionless kinetic equation. These components interact with each other through the gravitational potential $\phi$. Defining by $\mathbf{v}$ the velocity of dark matter particles with mass $m_d$ and through $\mathbf{u}$ the velocity of the gas with density $\rho$ and pressure $p$ we have the following equations:
\begin{align}
	&\frac{\partial f}{\partial t} + \mathbf{v}\frac{\partial f}{\partial \mathbf{r}} - \frac{\partial \phi}{\partial \mathbf{r}} \frac{\partial f}{\partial \mathbf{v}}  = 0, \label{kineticUP}\\
	&\frac{\partial \rho}{\partial t} +  \nabla {\rho \mathbf{u}} = 0,\\
	&\rho\frac{\partial \mathbf{u}}{\partial t} + \rho (\mathbf{u}\cdot \nabla)\mathbf{u} = -\nabla p - \rho\nabla \phi, \\
	& \Delta \phi = 4\pi G \left( \rho + m_d \int d\mathbf{v} f  \right)\label{potUP},
\end{align}
with ideal equation of state $p={kT}\rho/{m_b}$.

Let the initial configuration be stationary and consider small perturbations: $\rho = \rho_0(\mathbf{r}) + \rho_1(\mathbf{r},t)$, $p = p_0(\mathbf{r}) + p_1(\mathbf{r},t)$, $\mathbf{u} = 0+\mathbf{u}(\mathbf{r},t)$, $f = f_0(\mathbf{v},\mathbf{r}) + f_1(\mathbf{v},\mathbf{r},t)$, $\phi = \phi_0(\mathbf{r}) + \phi_1(\mathbf{r},t)$. Then the initial configuration is described with the following equations: 
\begin{align}
	&\nabla p_0 = -\rho_0\nabla\phi_0, \\
	&\mathbf{v}\frac{\partial f_0}{\partial\mathbf{r}} - \nabla{\phi_0} \frac{\partial f_0}{\partial\mathbf{v}} =0, \label{unprtbdF} \\
	&\Delta\phi_0 = 4\pi G \left( \rho_0 + m_d \int d\mathbf{v} f_0  \right), \\
	& p_0=c_s^2\rho_0,
\end{align}
where $c_s^2 = {kT}/{m_b}$ is the sound speed.

Equation~(\ref{unprtbdF}) has the formal solution $f_0(\mathbf{r},\mathbf{v})=F_0(\phi_0(\mathbf{r}) + \mathbf{v}^2/2)$ with an arbitrary $F_0$. Supposing that kinetic energy of DM particles is much more than gravitational energy (see Section~\ref{sec:longwave} for details) we have homogeneous spatial distribution: $f_0 = n_0 F(\mathbf{v})$, where $n_0=const$ is the DM particles concentration. 
We limited ourselves to considering such a case without assuming any coordinate dependence of the DM density.
Therefore the Jeans swindle should work for the DM component of the potential (see Section~\ref{sec:swindle}), i.e.\@ propagation of perturbations occurs in the presence of only BM initial potential. For the spherical symmetric case the BM unperturbed distribution is described in Section~\ref{sec:unprtbdIsoterm}. Since the initial configuration is stationary, we may set the perturbations to be proportional to $e^{-i\omega t}$, thus we were not interested in the initial conditions of the perturbations. Then the equations describing evolution of small perturbations are following:
\begin{align}
	&-i\omega f_1 +\mathbf{v}\frac{\partial f_1}{\partial\mathbf{r}} - \nabla{\phi_0} \frac{\partial f_1}{\partial\mathbf{v}} = n_0\frac{\partial F}{\partial \mathbf{v}}\nabla\phi_1 , \label{kineticPAT}\\
	&-i\omega \rho_1 +\nabla\rho_0\mathbf{u} = 0, \label{contPAT} \\
	&-i\omega \rho_0 \mathbf{u} = -c_s^2\nabla \rho_1 -\rho_0\nabla\phi_1 -\rho_1\nabla\phi_0 \label{velPAT} \\
	& \Delta \phi_1 = 4\pi G \left( \rho_1 + m_d \int d\mathbf{v} f_1  \right)\label{potPAT}.
\end{align}
The positive sign of the imaginary part of $\omega$ corresponds to an unstable initial configuration. Thus, solving these equations with the corresponding boundary conditions discussed below, and finding $\omega$ as an eigenvalue, one can make conclusion about the stability of an isothermal gas cloud in the presence of DM for given $m_b$, $T$, $P_s$ and $m_d$, $n_d$, $F(\mathbf{v})$.

\subsection{Longwave approximation} \label{sec:longwave}
Since the assumption has been made that the kinetic energy of the DM particles is much greater than the energy of the gravitational field, the third term of (\ref{kineticPAT}) can be neglected in comparison with the second one. Considering small increments for the corresponding terms one can conclude that this means that the work done by the gravitational field on a particle is small compared to the kinetic energy that the particle originally had. For interstellar gas clouds with their small gravitational gradients this approximation holds well (see also Section~\ref{sec:appLim}). 

Since a spatially endless DM distribution is considered, the localized gas distribution can be represented in zero approximation as
\begin{align}
	\rho_1^{(0)} = M_1\delta(\mathbf{r}), \,\,\, M_1=const, \label{zeroApprox}
\end{align}
and then (\ref{kineticPAT}) and (\ref{potPAT}) may be solved to continue with (\ref{contPAT})(\ref{velPAT})(\ref{potPAT}) with $f_1$ obtained in this way. 
Using Fourier transform it is easy to get the solution of (\ref{kineticPAT})(\ref{potPAT})(\ref{zeroApprox}):
\begin{align}
	& f_1 = 4\pi G n_0 M_1\int \frac{d\mathbf{k}} {(2\pi)^3} \frac{1}{\omega-\mathbf{k}\mathbf{v}} \frac{e^{i\mathbf{k}\mathbf{r}}}{k^2-\chi(\mathbf{k},\omega)}\mathbf{k}\frac{\partial F}{\partial\mathbf{v}} , \label{f1defen}
\end{align}
where 
\begin{align}
	\chi(\mathbf{k},\omega) = 4\pi G m_d n_0 \int  \frac{d\mathbf{v}}{\omega-\mathbf{k}\mathbf{v}}\mathbf{k}\frac{\partial F}{\partial\mathbf{v}} . \label{chiDef}
\end{align}
Defining the presenting in (\ref{potPAT}) integral as $\eta$ we have:
\begin{align}
	\eta(\mathbf{r}) = \frac{m_d}{M_1} \int d\mathbf{v}  f_1= \int \frac{d\mathbf{k}}{(2\pi)^3} \frac{\chi(\mathbf{k},\omega) e^{i\mathbf{k}\mathbf{r}}}{k^2-\chi(\mathbf{k},\omega)}, \label{etaDef}
\end{align}
then the (\ref{potPAT}) reads:
\begin{align}
	 \Delta \phi_1 = 4\pi G \left( \rho_1 + M_1 \eta(\mathbf{r}) \right). \label{potRHS}
\end{align}

Considering (\ref{contPAT})(\ref{velPAT})(\ref{potRHS}) we may see that they must have the same temporal behavior (the same increment $\omega$). Thus, the solution of these equations with appropriate boundary conditions will give the seeking increment $\omega$.

The initial assumption about localized BM density distribution implies that the propagating in DM medium perturbations have wave length which is larger than spatial scale of a gas cloud, i.e.\@ $k$ is small. Therefore (\ref{chiDef}) may be simplified:
\begin{align}
\chi(\mathbf{k},\omega) = -\frac{k^2}{\omega^2 \tau_d^2}\left( 1+ \frac{2}{\omega} \overline{(\mathbf{k}\cdot \mathbf{v})} + \frac{3}{\omega^2} \overline{(\mathbf{k}\cdot \mathbf{v})^2} \right) +\mathcal{O}(k^3),
\end{align} 
where the overline means the averaged over a distribution values and $\tau_d = (4\pi G m_d n_0)^{-1/2} $. We neglected the second term supposing the absence of a directed DM movement. Then, using (\ref{etaDef}) definition, $\eta(\mathbf{r})$ was gotten:
\begin{align}
	\eta(\mathbf{r}) = -\frac{1}{1+\omega^2\tau_d^2}\left(   \delta(\mathbf{r}) -\frac{3\tau_d^2}{1+\omega^2\tau_d^2} \overline{v_\alpha v_\beta}\frac{\partial^2  }{\partial r_\alpha r_\beta} \delta(\mathbf{r})  \right), \label{etaFin}
\end{align}  
where the presence of the delta-functions is a consequence of the assumption about the zero-order BM density spatial distribution (\ref{zeroApprox}). 

\subsection{Spherical symmetric case}\label{sec:spherSym}

\begin{figure}
	\centering
	\includegraphics[width=.71\linewidth]{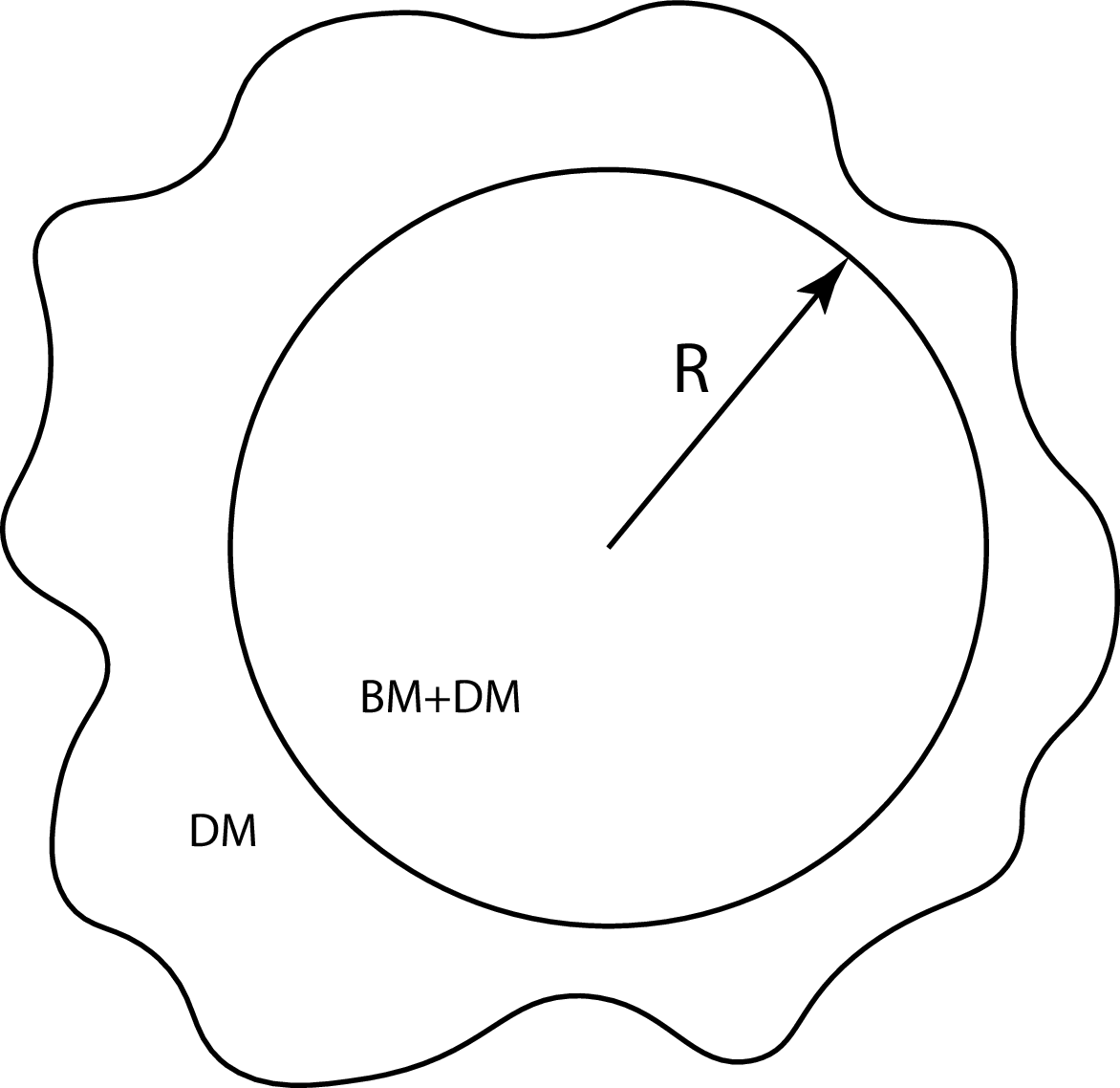}
	\caption{Representation of a spherical symmetric baryonic matter (BM) configuration in the presence of endless dark matter (DM).}
	\label{fig:DMandBM}
\end{figure}

 We restricted ourselves to consider only spherical symmetric case of initial gas configuration and perturbation propagation (see Figure~\ref{fig:DMandBM}). Therefore, following by \cite{Chavanis02b}, we introduced mass variables for $\rho_1$ and $\eta$ as follows:
\begin{align}
	& \rho_1 = \frac{1}{4\pi r^2}\frac{d q}{d r}, \label{massBM} \\
	& \eta = \frac{1}{4\pi r^2}\frac{d \Theta}{d r}.
\end{align}  
Then the perturbed Poisson equation (\ref{potRHS}) takes the just Newton's law form:
\begin{align}
	\frac{d\phi_1}{dr} = G\frac{q+M_1 \Theta}{r^2}, \label{pot1D}
\end{align}
which is a representation of the Gauss theorem. From continuity equation (\ref{contPAT}) we got:
\begin{align}
	u = \frac{i \omega}{4\pi r^2 \rho_0}q. \label{vel1D}
\end{align}
Then, after substituting (\ref{pot1D})(\ref{vel1D}) to (\ref{velPAT}) and some simple transformations the equation, describing the radial mass perturbation behavior, was gotten:
\begin{equation}
	\left( \frac{\omega^2}{4\pi \rho_0} + {G} \right) \frac{q}{r^2} +\frac{d}{dr} \frac{c_s^2}{4\pi r^2\rho_0} \frac{dq}{dr} = -G \frac{\Theta(r)}{r^2} M_1. \label{patGEN}
\end{equation}

This is the second order heterogeneous equation on eigenvalues $\omega$. Therefore the three boundary conditions are required -- two for determination arbitrary constants of homogeneous solutions and third for finding possible values of $\omega$. The Lagrangian derivative of perturbed pressure should equals to zero at the sphere bounder, which in our terms has the form:
\begin{align}
	\left( \frac{dq}{dr} - \frac{q}{\rho_0} \frac{d\rho_0}{dr} \right)_{r=R} =0. \label{bndConv}
\end{align}
Also, by definition (\ref{massBM}) we have:
\begin{align}
	q(0)=0, \label{bndZero}
\end{align}
which corresponds to zero mass including in zero radius sphere. 
Third heterogeneous boundary condition was gotten considering $M_1$ as full perturbed mass including in the sphere, so $M_1=4\pi\int \rho_1 r^2 dr$, which with (\ref{massBM}) leads to:
\begin{align}
	q(R) = M_1. \label{bndM1}
\end{align}

For spherically symmetric case the shape of $\Theta(r)$ 
is following:
\begin{align}
	\Theta(r) = -\frac{1}{1+\omega^2\tau_d^2}\left(  1 + \frac{3\tau_d^2 \overline{v^2}}{1+\omega^2\tau_d^2} \left(  \frac{2\delta(r)}{r} - \delta'(r)  \right)   \right), \label{zetaFin}
\end{align}
where the prime denotes $r$ derivative.

Thus, the formulated problem (\ref{patGEN})--(\ref{zetaFin}) completely determines the behavior of the perturbations and the stability of isothermal sphere in presents of DM with predefined BM density and DM velocity distributions.  This problem may be solved analytically for an arbitrary $\Theta(r)$ in case of constant density.  

\section{Stability investigation}
\subsection{Constant density}\label{sec:constDens}
Despite the fact that a stationary, spatially limited configuration configuration with $\rho_0,T=const$ cannot be realized (see also Section~\ref{sec:swindle}), we considered it because there is an exact solution for perturbations evolution in this case. It helped to construct a problem for isotherm distribution (see Section~\ref{sec:isothermStabPrblDet}) and reveal some quality characteristics which should be expected from consideration of realizable equilibrium configurations.

By setting $\rho_0=const$ we made $q(r)=r^{3/2}g(r)$ replacement which transforms (\ref{patGEN})--(\ref{bndM1}) problem to the form:
\begin{align}
	& r^2 g'' +r g' +\left( \nu^2 r^2 -\frac 9 4 \right)g = -l^{-2}r^{1/2}\Theta(r) M_1, \\
	& g(0)=0, \label{constDensB1}\\
	& R g'(R)+\frac{3}{2}g(R) =0, \label{constDensB2}\\
	& g(R) = R^{-3/2}M_1, \label{constDensB3}
\end{align}
where $l^{-2}=4\pi G\rho_0 c_s^{-2} = \tau_b^{-2} c_s^{-2}$ and $\nu^2 = c_s^{-2}\omega^2 + l^{-2}$. The general solution of homogeneous part is the combination of Bessel functions of $3/2$ and $-3/2$ orders therefore it is possible to set up solution of heterogeneous equation via Wronskian. After some transformations the general solution may be written as:
\begin{align}
	g(r) = &C_1 J_{3/2} (\nu r) + C_2 J_{-3/2} (\nu r) + \nonumber \\
	& + \frac{M_1}{l^2\nu^3}r^{-3/2} \int_{0}^{r} dz \,\, z^{-2}\Theta(z) \Big\{ \nu(r-z) \cos[\nu(r-z)] \Big. \nonumber \\
	&\Big. -  (1+\nu^2 rz) \sin[\nu(r-z)] \Big\}.
	\label{genSolConstDens}
\end{align}
With the boundary conditions  (\ref{constDensB1})(\ref{constDensB2})(\ref{constDensB3}) we came to the equation for the increment:
\begin{align}
\frac{R}{l^2 }\sqrt{\frac \pi 2} \int_{0}^{R} \frac{J_{3/2} (\nu z) \Theta(z)}{\sqrt{\nu z}} dz = \sin \nu R.
\end{align}
In the absent of DM (i.e. $\zeta=0$) the usual Jeans radius may be gotten. Substituting there (\ref{zetaFin}) and dividing it on $\nu R \neq 0$ we got the equation:
\begin{align}
	\frac{\sin\left(\varkappa\sqrt{1+\lambda^2}\right)}{\varkappa\sqrt{1+\lambda^2}} = -\sigma\frac{\sigma +\lambda^2  +\eta  (1+\lambda^2)}{\lambda^2(\sigma +\lambda^2)(1+\sigma+\lambda^2)}, 
	\label{constEq}
\end{align}
where $\lambda^2=\tau_b^2\omega^2$ -- dimensionless increment, $\varkappa = R/c_s\tau_b$, $\sigma= \tau_b^2/\tau_d^2 = m_d n_0/\rho_b$ -- ratio of the densities of DM and BM, $\eta = 3\overline{v^2}/c_s^2$ -- the ratio of the mean square speed of DM particles to the sound speed in gas.

\begin{figure}
	\centering
	\includegraphics[width=1\linewidth]{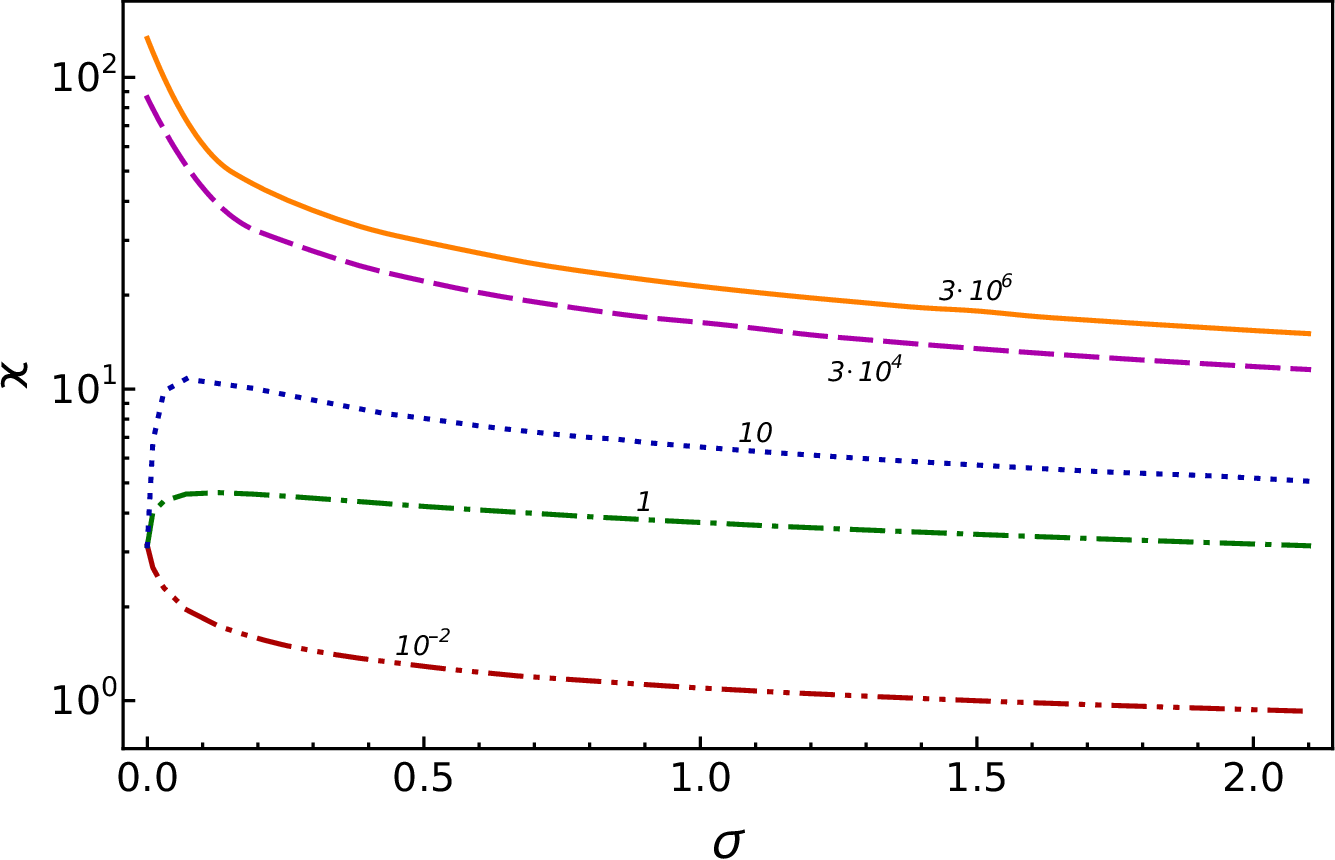}
	\caption{Dependence of the maximum stable dimensionless radius $\varkappa$ of an homogeneous sphere on the relative concentration of DM particles $\sigma = m_d n_0/\rho_0$ at various relative mean square velocities $\eta=3\overline{v^2}/c_s^2$:  $\,\, 3\cdot 10^6$ (solid line), $3\cdot 10^4$ (dashed), $10$ (dotted), $1$ (dot-dashed) and $10^{-2}$ (double-dot-dashed).
	}
	\label{fig:constDens}
\end{figure}

\begin{figure*}
	\begin{minipage}[h]{.49\linewidth}
		\center{\includegraphics[width=1\linewidth]{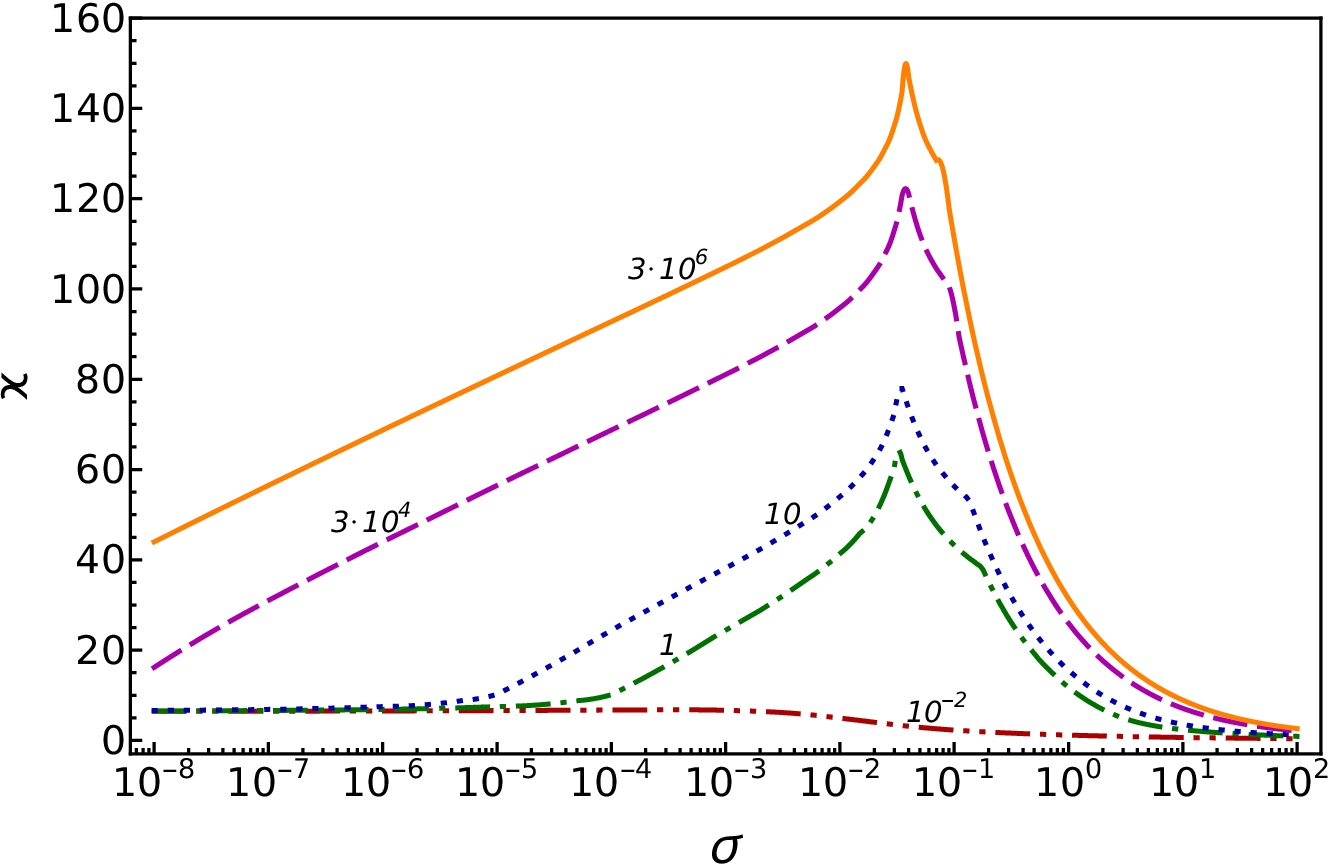}}
	\end{minipage}
	\hfill
	\begin{minipage}[h]{.49\linewidth}
		\center{\includegraphics[width=1\linewidth]{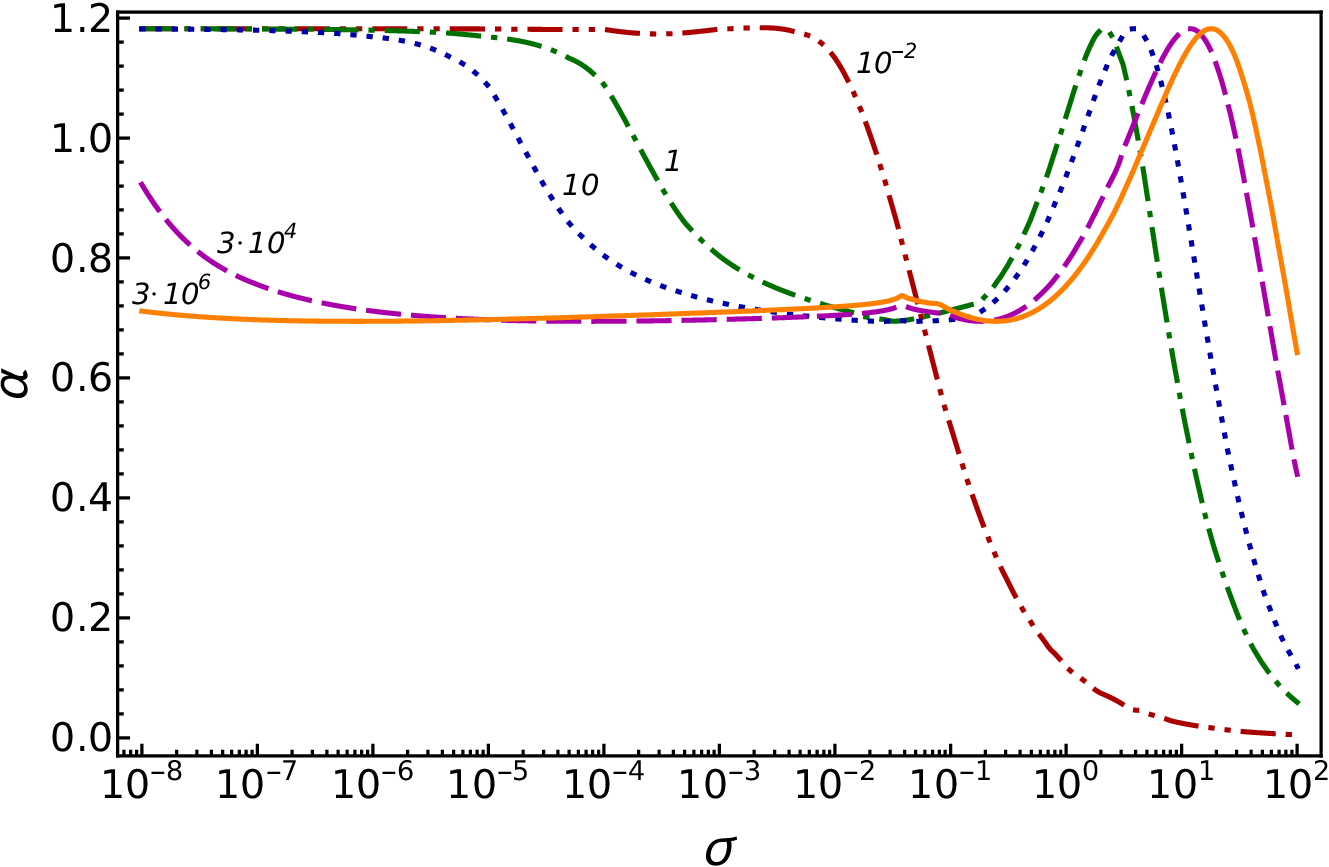}}
	\end{minipage}
	\vfill
	\smallskip\smallskip
	\begin{minipage}[h]{.49\linewidth}
		\center{\includegraphics[width=1\linewidth]{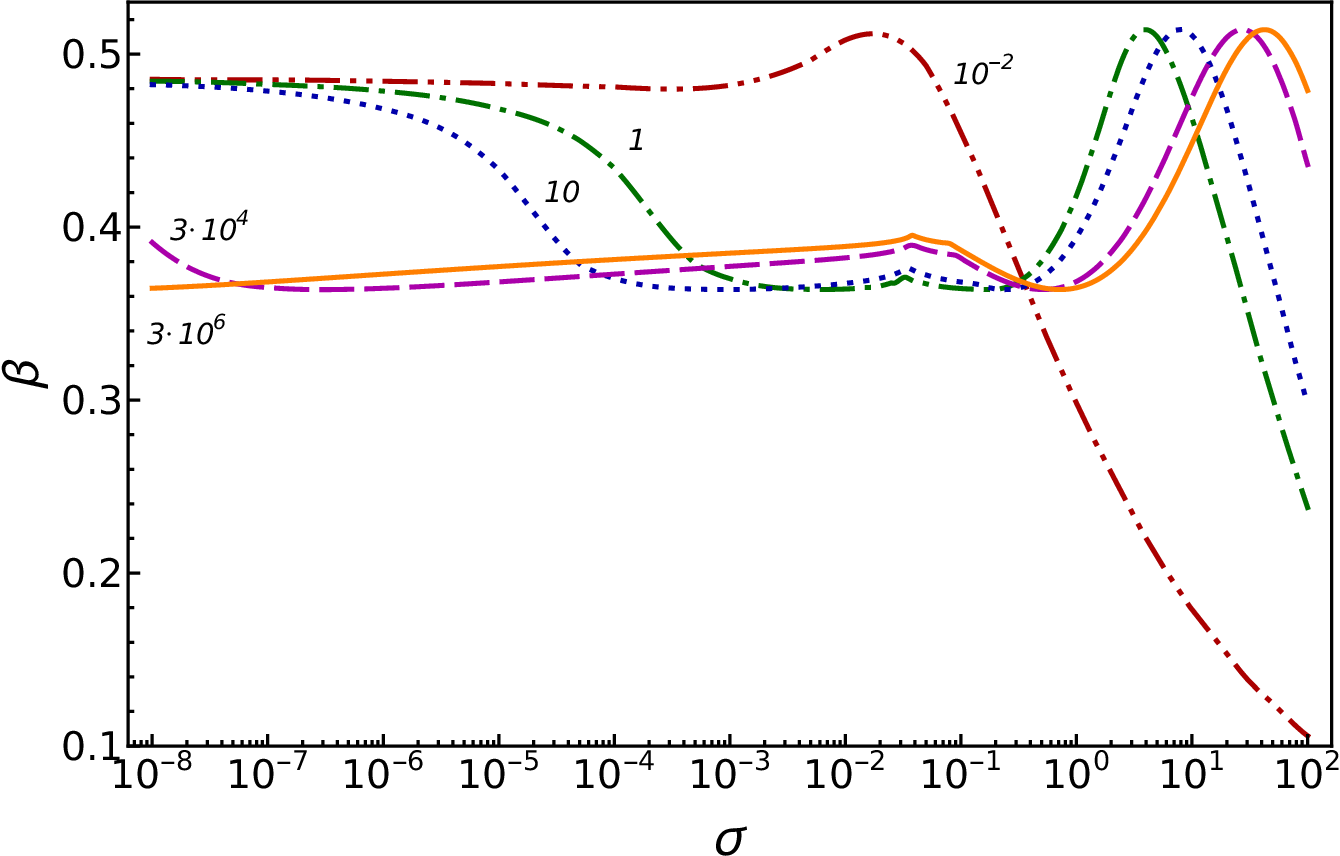}}
	\end{minipage}
	\hfill
	\begin{minipage}[h]{.49\linewidth}
		\center{\includegraphics[width=1\linewidth]{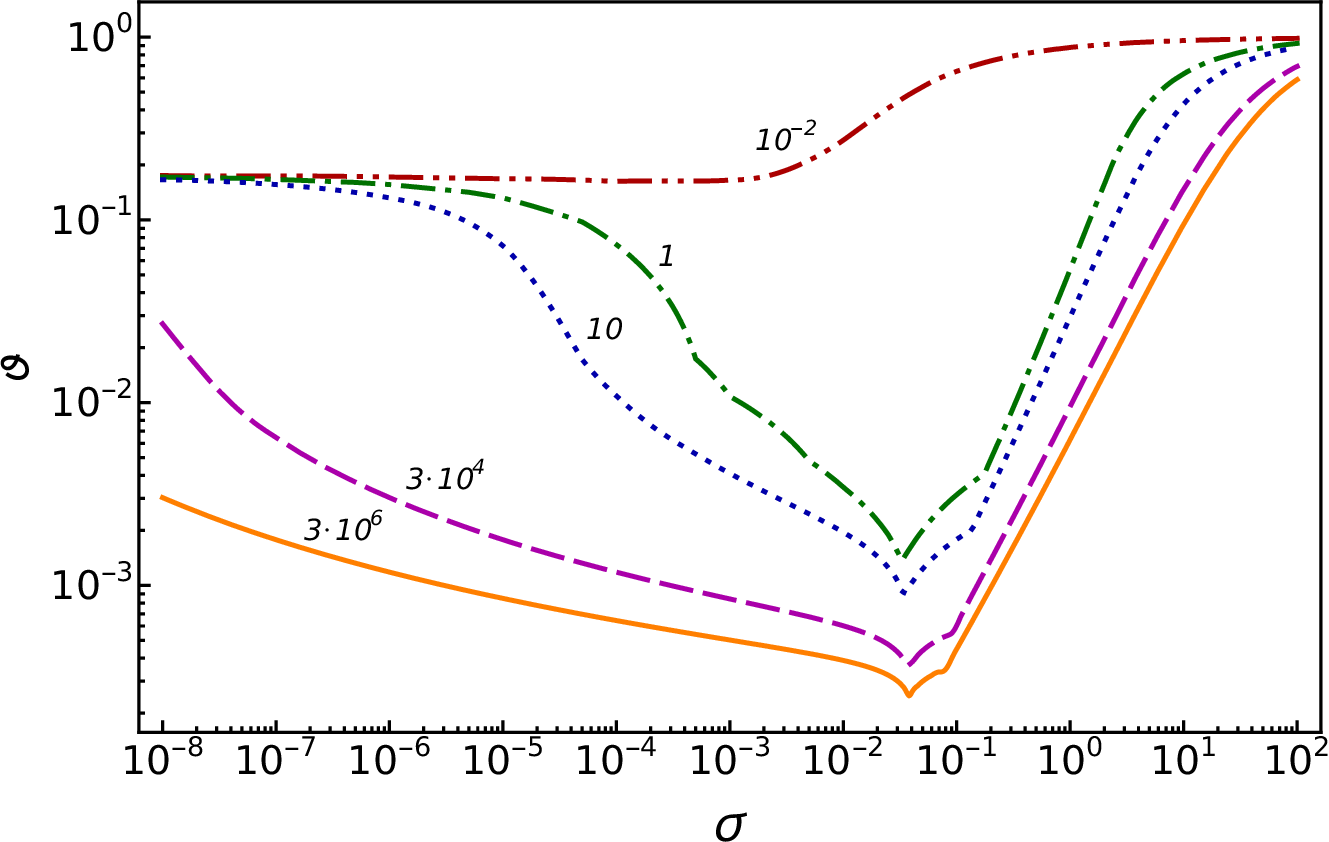}}
	\end{minipage}
	\caption{Dependence of the maximum stable dimensionless radius $\varkappa$ (top left panel), dimension radius ($\beta$ from equation~(\ref{BErad}), left bottom), mass ($\alpha$ from equation~(\ref{BEmass}),  top right) and ratio of average-to-central densities $\vartheta=\overline{\rho}/\rho_c$ (right bottom) of an isothermal sphere on the relative density of DM particles $\sigma = m_d n_0/\rho_c$ with various relative mean square velocities $\eta=3\overline{v^2}/c_s^2$:  $\,\, 3\cdot 10^6$ (solid line), $3\cdot 10^4$ (dashed), $10$ (dotted), $1$ (dot-dashed) and $10^{-2}$ (double-dot-dashed).	}
	\label{fig:isotermStab}
\end{figure*}

First of all it should be noted that, as follows from (\ref{constEq}), when DM particles are at rest (i.e.\@ $\eta=0$), any arbitrarily small portion of DM leads to the contraction of the gas sphere. This is because the resting gravitating DM particles cannot stay in equilibrium and begin to collapse, pulling along the gas, which interacts with them via the gravitational potential. However, this case does not apply to our consideration due to the assumption made in Sections~\ref{sec:govEqs},~\ref{sec:longwave} about the relatively large kinetic energy of the DM particles. To check how the mean square velocity affects the stability, several values of $\eta$ were chosen, for each of which the dependence of the critical radius $\varkappa$ on the relative density of DM particles $\sigma$ was calculated. 

The results of the calculations are presented in Figure~\ref{fig:constDens}, from which it can be seen that the small addition of slow DM particles leads to a increase in the radius $\varkappa$, and, accordingly, the mass, of a homogeneous gas cloud. Herewith, an increase of the DM particle velocity $\eta$ leads to a significant increase of the stable radius. Moreover, as expected, an increase in the relative density $\sigma$ of the DM leads to a decrease of $\varkappa$ at any values of the particle velocity. For large values of $\eta$ at low DM densities, extremely large values of the radius/mass of a homogeneous cloud were obtained. The physically meaningful case of an isothermal sphere need not have such large stable radius values, but we expected it to behave similarly. The whole case with constant density can be considered only as a toy model. Having dealt with the characteristic features of this case, we proceeded to study of the stability of isothermal spheres.

\subsection{Isothermal sphere stability}\label{sec:isothermStab}

\subsubsection{Problem determination}\label{sec:isothermStabPrblDet}
For the convenience of further numerical calculations, we transferred the values associated with the delta-function from the right side of (\ref{patGEN})(\ref{zetaFin}) to the boundary condition (\ref{bndZero}). To do this, we solved the problem on the interval $[0,\epsilon]$, where $0<\epsilon\rightarrow 0$, therefore the density can be considered being unchanged in this interval. The corresponding general solution was already obtained in the Section~\ref{sec:constDens}. To satisfy the boundary condition (\ref{bndZero}) in the solution we had to nullify $C_2$ in (\ref{genSolConstDens}), so it turns out that the solution is expressed only in terms of the integral, which can be easily calculated for $\epsilon\rightarrow 0$. Thus, we got an expression for $q(\epsilon)$ and may solve the problem on the interval $[\epsilon,R]$. Defining $\mu = q/M_1$ and dimensionless values according to Sections~\ref{sec:unprtbdIsoterm},~\ref{sec:constDens} the following problem was gotten:
\begin{align}
	&(\lambda^2 e^\psi +1 ) \mu + \zeta^2 \frac{d}{d\zeta} \frac{e^\psi}{\zeta^2} \frac{d\mu}{d\zeta} = \frac{\sigma}{\sigma+\lambda^2}, \label{dimlessEQ} \\
	& \mu(\epsilon) = -\frac{\sigma}{(\sigma+\lambda^2)^2}\eta, \,\,\,\,\, 0<\epsilon\rightarrow 0,  \label{dimlessBndEps} \\
	& \mu(\varkappa) = 1, \label{dimlessBndM1} \\
	& \frac{d\mu}{d\zeta}\Big|_{\varkappa} = -\frac{d\psi}{d\zeta}\Big|_{\varkappa}, \label{dimlessBndConv}
\end{align} 
which may be examined as an eigenvalue problem for the set of parameters $\varkappa$, $\sigma$, $\eta$. To test the obtained problem (\ref{dimlessEQ})--(\ref{dimlessBndConv}) we calculated $\lambda$ for some arbitrary parameters with $\psi=0$. The resulting $\lambda$ were the same as in the case of $\rho=const$ for any small enough $\epsilon$ like $10^{-5}$ or $10^{-10}$. Therefore, believing that the problem was set correctly, we proceeded to the stability analysis with $\psi(\zeta)$ obtained from the equations (\ref{psiEQ})(\ref{psiEQbnd}).

\subsubsection{Analysis of results}
 The results of the calculations are presented in Figure~\ref{fig:isotermStab} in terms of $\sigma = m_d n_0/\rho_c$, although instead of $\rho_c$ another gas characteristic calculated from the relations in Section~\ref{sec:unprtbdIsoterm} may be used. It can be seen that, as in the case of constant density, the presence of slow ($\eta=10^{-2}$) DM particles only leads to a decrease in both the dimensionless stable radius and mass of a gas cloud. Corresponding dimensional radius $\beta$ from equation~(\ref{BErad}) increases slightly up to $\sigma\approx 0.03$ and then decreases thereafter. It can be seen that for faster particles the dimensionless radius considerably increases up to $\sigma\approx 0.03$, after which it sharply decreases. At the same time, with increasing $\sigma$, the values of the dimensional radius decreases to $\beta=0.37$, after which, at DM dominant state $\sigma\approx 10$, they return to values corresponding to the absence of DM ($\beta=0.49$), and then rapidly decrease. The mass of the cloud $\alpha$ from (\ref{BEmass}) behaves similarly, dropping to $\alpha=0.64$ at $\sigma\lesssim 10$, i.e.\@ decreasing by $2.5$ times compared to a gas cloud without DM. Moreover, for very fast particles ($\eta=3\cdot 10^{6}$) this occurs at very low DM densities ($\sigma\sim 10^{-7}$).

The above means that presence of DM particles makes it possible to realize configurations of an isothermal self-gravitating gas sphere lying in inaccessible without DM areas. Indeed, as it can be seen from the Figure~\ref{fig:pressStab}, the critical equilibrium point shifts to the right beyond the global maximum of $\pi_s$ as $\sigma$ increases. The greater the relative speed $\eta$, the further along the curve the equilibrium point can move before it starts to move backwards at the value $\sigma\approx 0.03$. When the DM amount sufficiently exceeds the amount of BM ($\sigma\gtrsim 10$), only small, low-mass stationary configurations are possible, corresponding to the near-zero positions in the Figure~\ref{fig:pressStab}. As far as it can be seen from our numerical calculations, the point $\sigma\approx 0.03$ is not the same for all $\eta$ considered, but differs in the third digit. Its presence was quite expected (see Section~\ref{sec:constDens}), but the calculations above were needed to determine its value. 

Since at collapse the DM particles are carried along by the collapsing gas, the collapse of DM and BM must start simultaneously (\cite{Zeldovich80}). That is why an increase in velocity of DM particles prevents the beginning of collapse, while an increase in their concentration promotes it. At the regime with dominating DM the BM contribution is insignificant and stability of such configuration is described by Jeans theory applied to the homogeneous DM we consider. 

Since the dimension radius $\beta$ from (\ref{BErad}) has non-monotonic dependence on dimensionless one $\varkappa$ and the behavior of $\beta$ and the mass $\alpha$ are similar, it is difficult to judge about the structure of a stable configuration of an isothermal sphere. To clarify this, we calculated $\vartheta$ -- the ratio of average density to density in the center for the critical configurations using (\ref{centDens})(\ref{BEmass})(\ref{BErad}) relations:
\begin{align} 
 \vartheta=\frac{\overline{\rho}}{\rho_c} = 3\frac{\psi'(\varkappa)}{\varkappa},
\end{align}
which is represented at Figure~\ref{fig:isotermStab}. It can be seen that the presence of relatively small amounts of DM leads to the formation of a gas cloud with a high concentration of matter in the center and low one at the periphery. Moreover, $\vartheta$ can achieve values $\sim 10^{-3}$, i.e.\@ almost all gas is concentrated around the center of a sphere. 
And, as in the case of the quantities discussed earlier, a small number of very fast DM particles are sufficient to produce this effect. Then, as $\sigma$ increases so does $\vartheta$, tending to values equal to to $1$, which means that in the case of DM dominance the density of an isothermal sphere is homogeneous, although, as it was shown above, such configurations must be spatially small. 

Without aiming at a detailed analysis of ISM conditions, we estimated effects of DM presence on an isothermal gas sphere. We took the typical for solar neighborhood parameters: $\rho_c=5\cdot 10^{-22}\text{ g cm}^{-3}$, $T=50\text{ K}$, $m_b=2\cdot 10^{-24}\text{ g}$ with corresponding $c_s=8.3\cdot 10^4\text{ cm}^2 \text{ s}^{-2}$ and usual ones for DM consisting of WIMPs: $m_d n_0 = 5\cdot 10^{-25}\text{ g cm}^{-3}$, $(\overline{v^2})^{1/2}=6\cdot 10^{7} \text{ cm}^2 \text{ s}^{-2}$. The obtained results are presented in Table~\ref{tab:sampleParam}, which shows that the presence of DM increases the maximum stable mass of such a cloud by a factor of four, the radius by five, and decreases the average density by a factor of almost thirty. 

	\begin{table}
		\begin{tabular}{lccccc}
			\hline
				&  $\varkappa$	& Mass, $M_\odot$ & Radius, ly &Mean density,$\text{ g cm}^{-3}$  \\ \hline
			Without DM &  6.45 & $1.2\cdot 10^3$ & 19.6 & $8.8\cdot 10^{-23}$ \\
			With DM &  31.36 & $4.8\cdot 10^3$ & 95 & $3.2\cdot 10^{-24}$ \\ \hline
		\end{tabular}
	\caption{Values of an critical isothermal gas cloud at typical interstellar conditions parameters without and with DM having $\sigma=10^{-3}$ and $\eta=3\cdot 10^{6}$. }
	\label{tab:sampleParam}
	\end{table}

\subsubsection{Applicability limits}\label{sec:appLim}
To check the limits of applicability we calculated the work done by the gravitational field of a gas cloud on a DM particle to move it from its boundary to the center, which turned out to be $\psi(\varkappa)m_d c_s^2$. According to Section~\ref{sec:longwave} this work must be much less that the kinetic energy $m_d \overline{v^2}/2$, therefore we got the following criterion:
\begin{align}
	\eta\gg \psi(\varkappa).
\end{align} 
$\psi(\varkappa)$ is represented at Figure~\ref{fig:pressStab}, from which it can be seen that for $\eta=10^{-2},1$ this condition is only fulfilled at large values of $\sigma$ when the radius of the sphere $\varkappa$ is exceptionally small. But for large values of the relative velocity of DM particles $\eta$ this condition is well satisfied, which means that both the spatial DM density distribution and the third term in (\ref{kineticPAT}) might be neglected.

\section{Conclusions}

The stability of self-gravitating isothermal gas spheres in the presence of homogeneously distributed DM was analyzed. The behavior of the gas was described hydrodynamically, while the behavior of the DM particles was described using the kinetic equation. In describing the evolution of small perturbations of the distribution function, the gravitational effect of the gas cloud on these perturbations was neglected (the third term in (\ref{kineticPAT})), which, according to Section~\ref{sec:appLim}, is a good approximation for relatively fast particles. In addition, in Section~\ref{sec:longwave} we choose the BM spatial distribution in zero approximation as delta-function since it is a finite configuration in the background of infinite distributed DM. Then we obtained the problem (\ref{patGEN})--(\ref{zetaFin}) of finding first approximation of the perturbations and the instability increment for the spherically symmetric case. Further, having obtained an exact solution for the case of constant density, we formulated a numerically convenient problem (\ref{dimlessEQ})--(\ref{dimlessBndConv}) for isothermal spheres. The calculation results presented in Figure~\ref{fig:isotermStab} show that increasing DM density up to $0.03 \rho_c$, where $\rho_c$ is a central density, allows the inaccessible stable configurations located to the right of the global maximum in Figure~\ref{fig:pressStab} to be realized. Moreover, the greater the root mean square velocity of DM particles, the more distant from the global maximum configuration can be realized. A further increase in the DM density leads to a shift of a critical stable configuration in the opposite direction. At low DM densities, the mass of the gas in the cloud and its radius can increase several times, reducing its average density. Calculations performed for typical for solar neighbourhood parameters of ISM and DM consisting of WIMPs showed (see Table~\ref{tab:sampleParam}) that DM presence increases the maximum stable mass of isothermal cloud by a factor of four, the radius by five, and decreases the average density by a factor of thirty. It should be noted that parameters of both DM and ISM can differ widely both for different parts of a galaxy and for a considered cosmological epoch. 

Thus, it was shown that the presence of even a small number of fast DM particles leads to a significant change in the parameters and structure of isothermal gas clouds, making the realization of unstable configurations possible. We have considered the simplest model of isothermal clouds, which does not take into account many effects important for ISM (see Section~\ref{sec:Intro}). In order to get some accurate observational implications that can tell us something about the nature of DM, it seems necessary to take those effects into account. However, as follows from our simplified consideration, such effects should be quite significant. 

When hydrogen in the central part of a star is depleted, the nuclear energy release here ceases and passes to the upper layers surrounding the burnt-out core. As a result, an isothermal gas sphere appears in the center of the star, which is under pressure from the overlying non-isothermal layers. Such stars that left the main sequence are called red giants. If the star is massive enough ($M\gtrsim 2.5 M_\odot$) then its core consists of non-degenerate gas and may be described as in Section~\ref{sec:unprtbdIsoterm}. Our work has shown that the presence of even a small amount of fast DM particles interacting with the gas only gravitationally leads to a significant increase in the maximum possible mass and size of the isothermal sphere, which should be reflected in the structure of the red giant and its subsequent fate. Although in this case our consideration may not be entirely appropriate because the influence of the star's gravitational field on the initial DM spatial distribution and the development of the instabilities may be quite significant. However, the DM presence must have a significant influence on the entire evolution of stars, from the protostellar clouds to the compact objects formation. Several recent works (\cite{Raen21,Ellis21,Sagun21,Lopes21,Karkevandi22}) have focused on the effects of DM particle captured by stars and theirs possible observational manifestations. In addition, it would be useful to consider the gravitational stability of stars in the approach presented in this paper to find out the influence of DM on their fates.

\section*{Acknowledgements}
IK thanks G. Kolomiytsev for productive discussions.
The authors are grateful to A. Doroshkevich for useful final remarks.

\textsc{Make science, not war } \bcpeaceandlove

\bibliographystyle{mnras}
\bibliography{lib} 





\bsp	
\label{lastpage}
\end{document}